\documentclass[useAMS,usenatbib]{mn2e}
\usepackage{graphicx}
\usepackage{txfonts}

\title[The NS-calibration]
      {Abundance determination in H\,{\Large II} regions
      from spectra without the
      [O\,{\Large II}]$\lambda$3727+$\lambda$3729 line}

\author[L.S.Pilyugin, L.Mattsson]
       {L.S.~Pilyugin$^{1}$
        and L.~Mattsson$^{2,3}$  \\
     $^{1}$ Main Astronomical Observatory
            of National Academy of Sciences of Ukraine,
            27 Zabolotnogo str., 03680 Kiev, Ukraine \\
     $^{2}$ DARK Cosmology Centre, Niels Bohr Institute,
            University of Copenhagen, Juliane Maries Vej 30,
            DK-2100, Copenhagen \O, Denmark\\
     $^{3}$ Department of Physics and Astronomy, Uppsala University,
            Box 516,
            751 20 Uppsala, Sweden \\
             }

\date{Accepted 2010 November 2. Received 2010 October 16; in original form 2010 July  4}

\begin{document}

\maketitle

\begin{abstract}
We suggest an empirical calibration for determination of oxygen and
nitrogen abundances and electron temperature in H\,{\sc ii} regions where the
[O\,{\sc ii}]$\lambda$3727+$\lambda$3729 line ($R_2$) is not available.
The calibration is based on the strong emission lines of O$^{++}$,
N$^+$, and S$^+$  (NS calibration) and derived using the spectra of
H\,{\sc ii} regions with measured electron temperatures as calibration
datapoints. The NS calibration makes it possible to derive abundances for
H\,{\sc ii} regions in nearby galaxies from the SDSS spectra where $R_2$ line
is out of the measured wavelength range,
but can also be used for the oxygen and nitrogen abundances
determinations in any H\,{\sc ii} region independently whether the nebular 
oxygen line [O\,{\sc ii}]$\lambda$3727+$\lambda$3729 is available or not.
The NS calibration provides reliable oxygen and nitrogen
abundances for H\,{\sc ii} regions over the whole range of metallicities.
\end{abstract}


\section{Introduction}


The Sloan Digital Sky Survey (SDSS) \citep{yorketal2000}
provides a very large database of spectra
of individual H\,{\sc ii} regions in nearby galaxies and global emission-line
spectra of distant galaxies.
The wavelength range of the SDSS spectra is 3800 -- 9300\AA\ so that
for nearby galaxies with redshift $z$ $\la$ 0.023, the
[O\,{\sc ii}]$\lambda$3727+$\lambda$3729 emission line is out of that range.
The absence of this line prevents the use of the standard $T_e$ method 
where  the [O\,{\sc ii}]$\lambda$3727+$\lambda$3729 line is used 
for the determination of the contribution of ion O$^+$ to the total 
oxygen abundance in  H\,{\sc ii} region.  Since the calibrated relations 
between metallicity and the relative fluxes of strong oxygen lines
usually involve the [O\,{\sc ii}]$\lambda$3727+$\lambda$3729 line,
the absence of this line prevents the use of such relations to determine abundances.

\citet{kniazevetal2004} have suggested that this problem can be avoided
by using a modification of the standard $T_{e}$ method, based on the
intensity of the auroral line [O\,{\sc ii}]$\lambda$7320+$\lambda$7330
instead of the intensity of the [O\,{\sc ii}]$\lambda$3727+$\lambda$3729
nebular line. 
This modification of the standard $T_{e}$ method was used also by 
\citet{magrinietal2007} for abundance determinations in
 H\,{\sc ii} regions in the disk of the nearby spiral galaxy M~33.
Another way to overcome this problem has been suggested by
\citet{pilyuginthuan2007} based on the relation, called the $ff$-relation,
between the auroral and nebular oxygen line fluxes
\citep{pilyuginetal2006}. Using the $ff$-relation, the nebular
[O\,{\sc ii}]$\lambda$3727+$\lambda$3729  oxygen line flux can be
estimated from the measured  auroral [O\,{\sc iii}]$\lambda$4363 and nebular
[O\,{\sc iii}]$\lambda$4959+$\lambda$5007 oxygen line fluxes.
However, both these methods are useful only if the auroral
[O\,{\sc iii}]$\lambda$4363 oxygen line is detected in the SDSS spectra,
which is not the case for the majority objects.

Pioneering work by \citet{pageletal1979} and \citet{alloinetal1979} gave a method for
deriving abundances in H\,{\sc ii} regions in cases where direct measurement of
the electron temperature is not possible. They suggested that the
locations of H\,{\sc ii} regions in some emission-line diagrams can be
calibrated in terms of their oxygen abundances. This approach to abundance
determination in H\,{\sc ii} regions, usually referred to as the
``strong line method'' has been widely used.  Numerous relations have been
proposed to convert different emission-line ratios into metallicity or
temperature estimates
\citep[e.g.][]{dopitaevans1986,vilchezesteban1996,pilyugin2000,pilyugin2001,
pettinipagel2004,tremontietal2004,pilyuginthuan2005,
liangetal2006,stasinska2006,thuanetal2010}.
Comparisons between abundances derived from various calibrations are
given in the recent papers of \citet{kewleyellison2008,lopezsanchez2010}.
It was found that the calibrations based on the strong nitrogen line provide
acceptable results for objects with 12+log(O/H)$\ga$8.0 only
\citep{perezmontero2009,lopezsanchez2010}.

It has been argued that the ratio of the nebular nitrogen line
[N\,{\sc ii}]$\lambda$6548+$\lambda$6584 flux to the nebular oxygen line
[O\,{\sc ii}]$\lambda$3727+$\lambda$3729 flux  and the ratio of the nebular sulfur
line [S\,{\sc ii}]$\lambda$6717+$\lambda$6731 flux to the nebular oxygen line
[O\,{\sc ii}]$\lambda$3727+$\lambda$3729 flux can be used as a surrogate
indicators of the electron temperature and the metallicity
\citep{pilyuginetal2009,pilyuginetal2010}. Based on this, a new
improved empirical calibration for the determination of electron
temperatures and oxygen and nitrogen abundances in H\,{\sc ii} regions
from the strong emission lines of O$^{++}$, O$^{+}$, N$^+$ and S$^+$
has recently been suggested \citep{pilyuginetal2010}.
This calibration provides reliable abundances for H\,{\sc ii}
regions over the whole range of metallicity. However,
the calibration relations involve the nebular oxygen line
[O\,{\sc ii}]$\lambda$3727+$\lambda$3729.

From a general point of view, one may expect the ratio of the nebular
nitrogen line [N\,{\sc ii}]$\lambda$6548+$\lambda$6584 flux to the nebular
sulfur line [S\,{\sc ii}]$\lambda$6717+$\lambda$6731 flux to be useful as
a surrogate indicator of the electron temperature and the metallicity in
H\,{\sc ii} region. If this is the case, then one can construct a calibration
for the determination of abundances in H\,{\sc ii} regions such that
the calibration relations will not involve the line
[O\,{\sc ii}]$\lambda$3727+$\lambda$3729. This is the main objective of our study and
we will follow the strategy suggested by \citet{pilyuginetal2010}.

Throughout the paper, we will be using the following notations for the line
fluxes,
\begin{equation}
R_2 = [{\rm O}\,\textsc{ii}] \lambda 3727+ \lambda 3729
    = I_{\rm [OII] \lambda 3727+ \lambda 3729} /I_{{\rm H}\beta },
\end{equation}
\begin{equation}
N_2 = [{\rm N}\,\textsc{ii}] \lambda 6548+ \lambda 6584
    = I_{\rm [NII] \lambda 6548+ \lambda 6584} /I_{{\rm H}\beta },
\end{equation}
\begin{equation}
S_2 = [{\rm S}\,\textsc{ii}] \lambda 6717+ \lambda 6731
    = I_{\rm [SII] \lambda 6717 + \lambda 6731} /I_{{\rm H}\beta },
\end{equation}
\begin{equation}
R_3 = [{\rm O}\,\textsc{iii}] \lambda 4959+ \lambda 5007
    = I_{{\rm [OIII]} \lambda 4959 + \lambda 5007} /I_{{\rm H}\beta }.
\end{equation}
\begin{equation}
R_{23} = R_2 + R_3 . 
\end{equation}
The electron temperatures $t$ are given in units of 10$^4$K.
Furthermore, we will be using the following terminology. We refer to a 'method' as
the basic principle for deriving an abundance, e.g., $T_{\rm e}$-method, strong line method.
'Calibration' refers to a specific form of the strong line method, e.g., the R23,
N2 or NS-calibration. We also refer to 'relations' as the resultant expressions for
determination of, e.g., $t$ or O/H.

\section{The NS-calibration}

Here we search for a calibration based on the strong emission lines of
O$^{++}$, N$^+$, and S$^+$. Since the ratio of the nebular nitrogen line
[N\,{\sc ii}]$\lambda$6548+$\lambda$6584 flux to the nebular sulfur line
[S\,{\sc ii}]$\lambda$6717+$\lambda$6731 flux will be used
as a surrogate indicator for electron temperature and
metallicity, the sought calibration will be referred to as the NS-calibration.

For the calibration, we have used a sample of 118 H\,{\sc ii} regions with metallicities
spanning a large range from \citet{pilyuginetal2010}.
In previous study it was found that none of the considered emission line fluxes
and flux ratios display a monotonic behavior with electron temperature and
oxygen and nitrogen abundances over the whole temperature (or metallicity)
range \citep{pilyuginetal2010}.  Instead, the curves show bends and
it is therefore preferable to construct relations for
electron temperatures and abundances for three distinct regimes.

\begin{figure}
\resizebox{1.00\hsize}{!}{\includegraphics[angle=000]{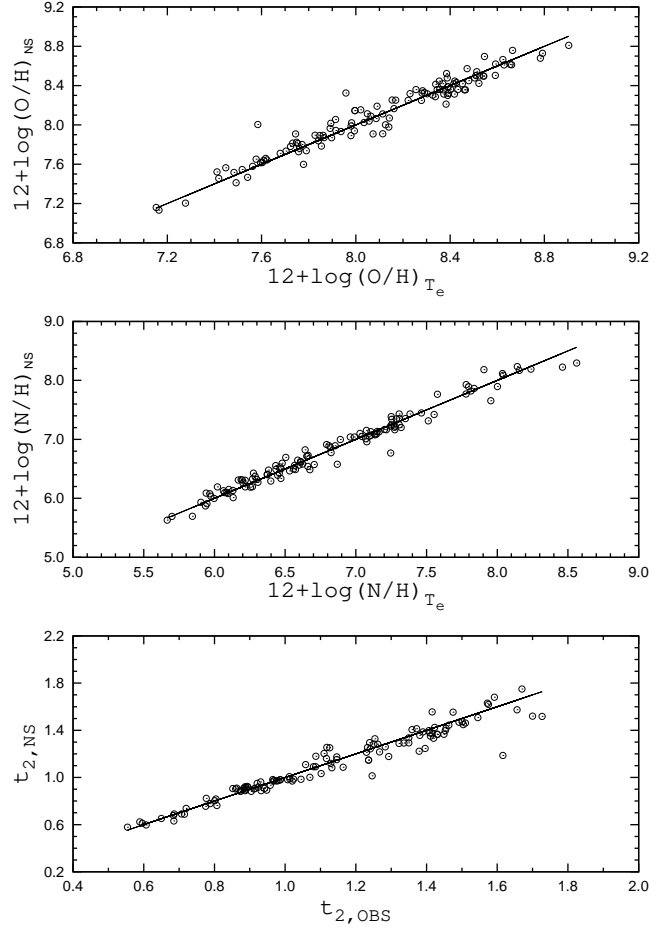}}
\caption{
The oxygen abundance (O/H)$_{\rm NS}$ (upper panel), nitrogen abundance
(N/H)$_{\rm NS}$ (middle panel), and electron temperature
$t_{\rm 2,NS}$ (lower panel) derived using the NS-calibration
against the measured $T_{\rm e}$-method values. The open circles
show the calibration H\,{\sc ii} regions, while the line shows the case of
equal values.
}
\label{figure:zztt}
\end{figure}

The following simple expression is adopted to relate the oxygen abundance to the
strong line fluxes,
\begin{equation}
12+\log {\rm (O/H)}  = a_0 +  a_1\,\log R_3 + a_2\,\log N_2 + a_3\,\log (N_2/S_2).
\label{equation:z}
\end{equation}
In cool high-metallicity H\,{\sc ii} regions, the $N_2$ line fluxes remain
approximately constant with temperature, forming a sort of plateau  \citep{pilyuginetal2010}. 
Therefore, we use log$S_2$ instead of log$N_2$ for cool high-metallicity
(with log$N_2$$>$--0.1) H\,{\sc ii} regions. The numerical values of the
coefficients in Eq.(\ref{equation:z}) can be derived requiring
that the scatter of the residuals $\sigma _{\rm O/H}$, defined as
\begin{equation}
\sigma _{\rm O/H}   = \sqrt{\frac{1}{n}\sum\limits_{j=1}^{j=n}
(\log {\rm (O/H)}_{j}^{CAL} - \log {\rm (O/H)}_{j}^{OBS})^2},
\label{equation:sig-z}
\end{equation}
is to be minimised. In the equation above, (O/H)$_j^{CAL}$ is the oxygen abundance
calculated from Eq.(\ref{equation:z}) and (O/H)$_j^{OBS}$ is the measured
$T_{\rm e}$-based oxygen abundance.

We derive the following relations for
determination of the oxygen abundance $Z_{\rm O}$=12+log(O/H)$_{\rm NS}$
\begin{eqnarray}
       \begin{array}{lll}
Z_{\rm O} & =   &  8.454 - 0.216\,\log R_3 - 0.362\,\log S_2-     \\
          &     &  0.101\,\log (N_2/S_2),                             \\
          &     & {\rm for} \quad\log N_2 > -0.1                        \\
          &     &                                                    \\
Z_{\rm O} & =   &  8.456 + 0.082\,\log R_3 + 0.391\,\log N_2+     \\
          &     &  0.290\,\log (N_2/S_2),                             \\
          &     & {\rm for} \quad\log N_2 < -0.1, \; \log (N_2/S_2) > -0.25      \\
          &     &                                                    \\
Z_{\rm O} & =   &  7.881 + 0.929\,\log R_3 + 0.650\,\log N_2+     \\
          &     &  0.025\,\log (N_2/S_2),                             \\
          &     &{\rm for} \quad\log N_2 < -0.1, \; \log (N_2/S_2) < -0.25.   \\
     \end{array}
\label{equation:oh}
\end{eqnarray}
A few datapoints deviate significantly ($>$0.3 dex) from the general trend. These were excluded
in the derivation of the final calibration relations, in order to avoid any bias.
The upper panel of Fig.~\ref{figure:zztt} shows the oxygen abundance
12+log(O/H)$_{\rm NS}$
derived using the NS-calibration against the $T_{\rm e}$-based oxygen abundance
12+log(O/H)$_{\rm T_e}$. The open circles show the calibration H\,{\sc ii}
regions. The line shows the case of equal values.
Fig.~\ref{figure:zztt} shows that the derived NS-calibration
provides reliable oxygen abundances in H\,{\sc ii} regions with only a
few exceptions.
The scatter of the residuals is $\sigma _{\rm O/H}$ = 0.077, excluding the two points
with largest deviations.

In the same way, a relation for nitrogen abundance determination
in H\,{\sc ii} regions has been constructed. The relations for nitrogen abundances
$Z_{\rm N}$=12+log(O/H)$_{\rm NS}$ are
\begin{eqnarray}
       \begin{array}{lll}
Z_{\rm N}  & =   &  7.414 - 0.383\,\log R_3 + 0.119\,\log S_2+      \\
           &     &  0.988\,\log (N_2/S_2),                              \\
           &     &  {\rm for } \quad \log N_2 > -0.1                                  \\
           &     &                                                     \\
Z_{\rm N}  & =   &  7.250 + 0.078\,\log R_3 + 0.529\,\log N_2+      \\
           &     &  0.906\,\log (N_2/S_2),                              \\
           &     &  {\rm for } \quad \log N_2 < -0.1, \; \log (N_2/S_2) > -0.25       \\
           &     &                                                     \\
Z_{\rm N}  & =   &  6.599 + 0.888\,\log R_3 + 0.663\,\log N_2+      \\
           &     &  0.371\,\log (N_2/S_2),                              \\
           &     &  {\rm for } \quad \log N_2 < -0.1, \; \log (N_2/S_2) < -0.25       \\
     \end{array}
\label{equation:nh}
\end{eqnarray}

The middle panel in Fig.~\ref{figure:zztt} shows a comparison between the nitrogen abundances
(N/H)$_{\rm NS}$ derived using the NS-calibration and the measured
abundances (N/H)$_{\rm T_e}$. The open circles show the calibration
H\,{\sc ii} regions, while the line shows the case of equal values.
Again, the NS-calibration provides reliable abundances also for nitrogen with only a few exceptions.
The scatter of the residuals is $\sigma _{\rm N/H}$ = 0.110, excluding the two points
with largest deviations of O/H.

Within the framework of the H\,{\sc ii} region model adopted in the
present study, the electron temperature is characterized by
two values $t_3$ and $t_2$. The value $t_2$ is the characteristic electron 
temperature of the O$^+$, N$^+$ zones, and the value $t_3$ is that of the O$^{++}$ zone. 
We look for a calibration to
determine $t_2$, which for brevity will referred to hereafter as simply $t$.
In the case where a measurement of $t_3$ is available for a calibration
H\,{\sc ii} region, $t_2$ is derived from the $t_2$ -- $t_3$ relation
\citep{pilyuginetal2009}. 
We adopt the following expression to relate the electron temperature to
the strong line fluxes,
\begin{equation}
t^{-1}  = {a_0 + a_1\,\log R_3 + a_2\,\log N_2 + a_3\,\log (N_2/S_2)}.
\label{equation:tg}
\end{equation}
Again, the numerical values of the coefficients in  Eq.(\ref{equation:tg})
are derived by minimisation of the scatter $\sigma _t$
\begin{equation}
\sigma _t   = \sqrt{\frac{1}{n}\sum\limits_{j=1}^{j=n}
\left(\frac{t_{j}^{CAL}}{t_{j}^{OBS}}-1\right)^2}.
\label{equation:sig-t}
\end{equation}
In the equation above $t_j^{CAL}$ is the electron temperature $t_2$
calculated from Eq.(\ref{equation:tg}) and $t_j^{OBS}$ is the measured
$t_2$. We derive the following relations for the electron temperature
$t$ = $t_{\rm 2,NS}$
\begin{eqnarray}
       \begin{array}{lll}
t^{-1} & =   &  1.185 - 0.351\,\log R_3 - 0.273\,\log S_2 +  \\
       &     &  0.059\,\log (N_2/S_2),                           \\
       &     &  {\rm for } \quad \log N_2 > -0.1               \\
       &     &                                                   \\
t^{-1} & =   &  1.226 - 0.219\,\log R_3 + 0.133\,\log N_2 +             \\
       &     &  0.225\,\log (N_2/S_2),                                      \\
       &     &  {\rm for } \quad \log N_2 < -0.1, \; \log (N_2/S_2) > -0.25 \\
       &     &                                                              \\
t^{-1} & =   &  0.953 + 0.117\,\log R_3 + 0.230\,\log N_2 +               \\
       &     &  0.033\,\log (N_2/S_2),                                        \\
       &     &  {\rm for } \quad \log N_2 < -0.1, \; \log (N_2/S_2) < -0.25 \\
     \end{array}
\label{equation:t}
\end{eqnarray}
Since there is a correspondence between $t_2$ and $t_3$, the
same indicators can be used to determine both $t_2$ and $t_3$.
In other words, a similar calibration for $t_3$ can be also constructed.

In the lower panel in Fig.~\ref{figure:zztt} we plot the electron temperature
$t_{\rm 2,NS}$
derived using the NS-calibration against the measured electron temperature
$t_{\rm 2,OBS}$. As before, the open circles show the calibration H\,{\sc ii} regions,
while the line shows the case of equal values.
Fig.~\ref{figure:zztt} shows that the values of the electron
temperature derived using the NS-calibration are close to the measured
temperatures in H\,{\sc ii} regions with only a few exceptions.
The scatter of the residuals is $\sigma _t$ = 0.058, excluding the two points
with largest deviations of O/H.

\section{Discussion}

\begin{figure}
\resizebox{1.00\hsize}{!}{\includegraphics[angle=000]{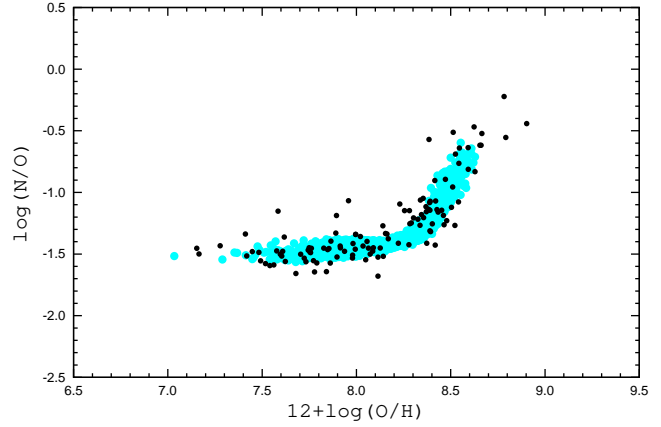}}
\caption{
The O/H--N/O diagram.
The filled gray (light-blue in the color version) circles show abundances
derived using the NS-calibration for the sample of nearby SDSS galaxies.
The filled black circles show
$T_e$-based abundances in our sample of calibration H\,{\sc ii}
regions.
(A color version of this figure is available in the online version.)
}
\label{figure:ohno}
\end{figure}

To verify the reliability of the NS-calibration,
we compare the N/O--O/H diagram
using the NS-calibration for SDSS spectra of H\,{\sc ii} regions
with that obtained from H\,{\sc ii} regions in nearby galaxies with $T_e$-based
abundances (the sample of calibration H\,{\sc ii} regions).
Line flux measurements in SDSS spectra have been carried out by several
groups. Here, we use the data catalogs made publicly available
by the MPA/JHU group \footnote{The catalogs are available at
http://www.mpa-garching.mpg.de/SDSS/}.
The techniques used to construct the catalogues are described
in \citet{brinchmannetal2004,tremontietal2004}
and other publications by the same authors. We have chosen to use these
catalogues instead of the original SDSS spectral database because the line flux
measurements are generally more accurate.

We extracted emission-line objects from the MPA/JHU catalogs
which satisfy the following criteria:
(1) the redshift z$\la$0.023,
(2) the equivalent widths of
H$\alpha$,
H$\beta$,
[O\,{\sc iii}]$\lambda$4959,
[O\,{\sc iii}]$\lambda$5007,
[N\,{\sc ii}]$\lambda$6584,
[S\,{\sc ii}]$\lambda$6717 and
[S\,{\sc ii}]$\lambda$6731 lines are larger than 3\AA,
3) the line ratio
3.3$\ga$[O\,{\sc iii}]$\lambda$5007/[O\,{\sc iii}]$\lambda$4959$\ga$2.7.
Since our calibrations are valid only in the low-density regime, we have
just included objects with a reasonable value of the [S\,{\sc ii}]
line ratio, i.e., those with
1.1 $<$ F$_{\rm [S\,II]\lambda 6717}$/F$_{\rm [S\,II]\lambda 6731}$ $<$ 1.6.
Applying these criteria, we selected 4339 spectra
from the MPA/JHU catalogs.
Measurements of the [N\,{\sc ii}]$\lambda$6584 line are more reliable than
those of the [N\,{\sc ii}]$\lambda$6548 line. Hence, we have used
N$_2$ = 1.33[N\,{\sc ii}]$\lambda$6584 instead of the standard
N$_2$ = [N\,{\sc ii}]$\lambda$6548 + [N\,{\sc ii}]$\lambda$6584.
The emission fluxes are then corrected for interstellar reddening
using the theoretical H$\alpha$/H$\beta$-ratio and the analytical
approximation to the Whitford interstellar reddening law from
\citet{izotovetal1994}. Occationally, the derived
values of the extinction c(H$\beta$) are negative (but close to zero).
In those cases we set c(H$\beta$)$=0$.

We have derived (O/H)$_{NS}$ and (N/H)$_{NS}$  for all H\,{\sc ii} regions
in the extracted SDSS sample. Fig.~\ref{figure:ohno} shows the resultant O/H--N/O diagram.
SDSS H\,{\sc ii} regions are plotted with filled gray (light-blue in the color version) circles.
For comparison, H\,{\sc ii} regions in nearby galaxies with $T_{\rm e}$-based
abundances (the sample of calibration H\,{\sc ii} regions)
are shown as filled black circles.
Fig.~\ref{figure:ohno} also shows that the
SDSS H\,{\sc ii} regions nicely follow the general trend in the N/O--O/H diagram
traced by the H\,{\sc ii} regions with $T_{\rm e}$-based abundances.

\begin{figure}
\resizebox{1.00\hsize}{!}{\includegraphics[angle=000]{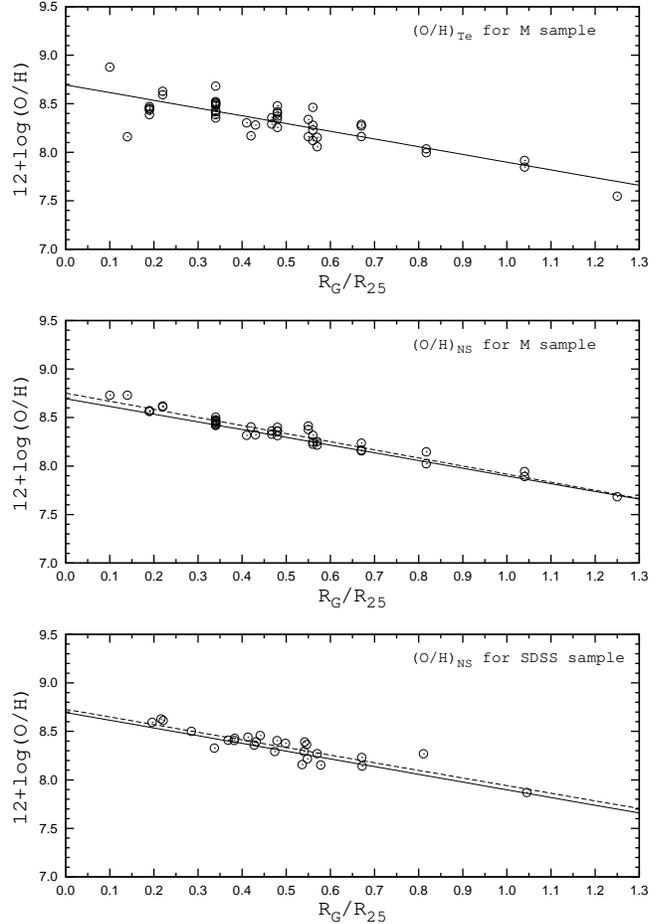}}
\caption{
Radial distribution of oxygen abundances in the disk of the spiral galaxy
M~101 determined in different ways.
{\it Upper panel:} Distribution of (O/H)$_{T_e}$ abundances
for the $M$ sample (see text).
The solid line is the best fit to the data.
{\it Middle panel:} Distribution of (O/H)$_{\rm NS}$ abundances
for the $M$ sample.
The dashed line is the best fit to these data, the solid line is the same as in the
upper panel.
{\it Lower panel:} Distribution of (O/H)$_{\rm NS}$ abundances
based on SDSS spectra.
The dashed line is the best fit to these data,
the solid line is the same as in the upper panel.
}
\label{figure:m101oh}
\end{figure}

\begin{figure}
\resizebox{1.00\hsize}{!}{\includegraphics[angle=000]{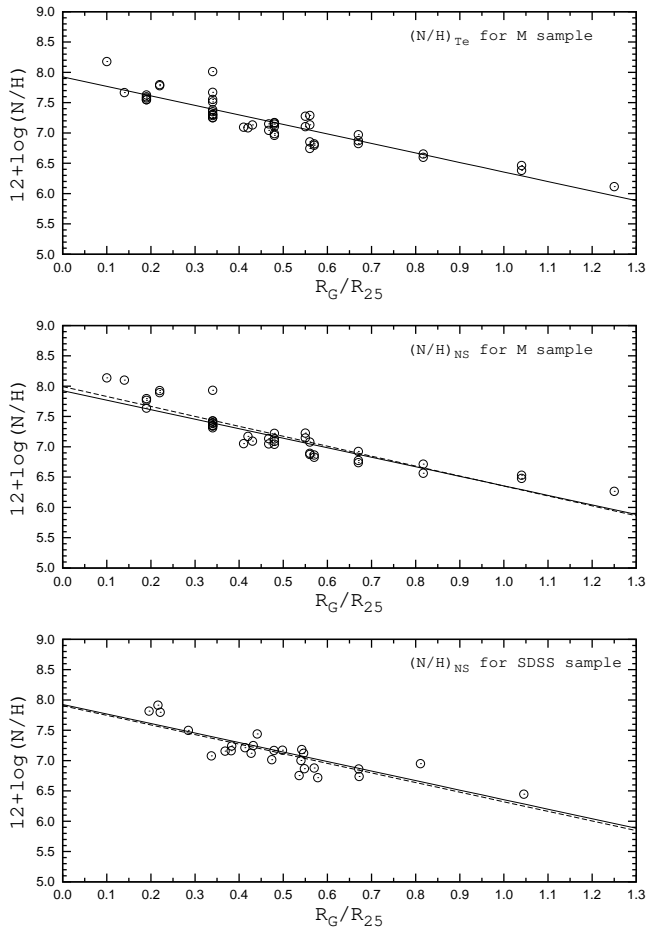}}
\caption{
The same as in Fig.~\ref{figure:m101oh} but for nitrogen.
}
\label{figure:m101nh}
\end{figure}

We have also tested the reliability of the NS-calibration using data for the spiral
galaxy M~101 (=NGC~5457) as a reference.
The radial distributions of oxygen and nitrogen abundances in the
disk of M~101 are well established by the $T_e$-method, and
the oxygen abundances in H\,{\sc ii} regions in
M~101 span a large interval of metallicity (about an order of magnitude),
which provides a possibility to test the NS-calibration in
different metallicity regimes.
Furthermore, there are spectra for several dozens of H\,{\sc ii} regions in
M~101 in the SDSS spectral database.

The spectra of H\,{\sc ii} regions in the disk of
M~101 with measured electron temperatures $t_3$=$t$(O\,{\sc iii})
or/and $t_2$=$t$(N\,{\sc ii}) have been taken from
\citet{sedwickaller1981,rayoetal1982,mccalletal1985,torresetal1989,
garnettkennicutt1994,kinkelrosa1994,vanzeeetal1998,luridianaetal2002,
kennicuttetal2003,izotovetal2007,estebanetal2009}. For the sake of brevity,
this compilation of spectra will be referred to as the $M$ sample.
The $M$ sample contains 47 measurements of electron temperature. 
It should be noted that a significant fraction (30 out of 47) of the $M$ sample is
part of the sample of the calibration data points.
The deprojected galactocentric distances of H\,{\sc ii} regions normalised to
the disk isophotal radius are taken from
\citet{kennicuttgarnett1996,kennicuttetal2003}.

Fig.~\ref{figure:m101oh} shows the radial distributions of oxygen abundances
in the disk of M~101 determined in different ways. The upper panel shows the
distribution of (O/H)$_{T_e}$ abundances for the $M$ sample.
The solid line is the linear least-square best fit to the data.
The middle panel in Fig.~\ref{figure:m101oh} shows the radial distribution
of (O/H)$_{\rm NS}$ abundances for the same $M$ sample.
The dashed line shows a linear least-square fit to these data.
The solid line is the same as in the upper panel.
One can see that the scatter of  (O/H)$_{\rm NS}$ abundances in the middle 
panel is lower respect to the scatter of  (O/H)$_{T_e}$ abundances in the 
upper panel. This may be evidence for that the larger scatter in the (O/H)$_{T_e}$
abundances is caused, at least partly, by the uncertainties in the measurements 
of weak auroral lines.

The lower panel of Fig.~\ref{figure:m101oh}
shows the radial distribution of (O/H)$_{\rm NS}$ abundances
for the sample of selected SDSS spectra of H\,{\sc ii} regions in the disk of the
M~101 (line flux measurements in SDSS spectra have been taken from
the MPA/JHU catalog using the selection criteria mentioned above).
After correction for interstellar reddening,
the oxygen (O/H)$_{\rm NS}$ and nitrogen (N/H)$_{\rm NS}$ abundances were
determined using Eqs.(\ref{equation:oh}),(\ref{equation:nh}).
The galactocentric distances of the SDSS H\,{\sc ii} regions have been
computed using the SDSS coordinates with a major
axis position angle of 37\degr and an inclination between the line of sight
and polar axis of 18\degr \citep{kennicuttgarnett1996}. The fractional
distances R$_{\rm G}$/R$_{25}$ have deen obtained with
the isophotal radius R$_{25}$ = 14.4 arcmin taken from
the Third Reference Catalog of bright Galaxies \citep{rc3}.
The dashed line shows a linear least-square fit to these data.
The solid line is the same as  in the upper panel.

Fig.~\ref{figure:m101oh} shows that the oxygen abundances
derived using the NS calibration for both the compilation of spectra
and the SDSS spectra follow the radial gradient traced by the
H\,{\sc ii} regions with oxygen abundances determined with the direct method.

Fig.~\ref{figure:m101nh} shows the radial distributions of nitrogen abundances
in the disk of M~101 determined in different ways.
Again the upper panel shows the distribution of (N/H)$_{T_e}$ abundances for
the $M$ sample, the middle panel shows the radial distribution
of (N/H)$_{\rm NS}$ abundances for the $M$ sample,
and the lower panel shows the radial distribution of (N/H)$_{\rm NS}$
abundances for a sample of SDSS spectra of H\,{\sc ii} regions in the disk
of M~101. Again, the nitrogen abundances
derived using the NS-calibration for both our compilation of spectra
and the SDSS spectra, follow the radial gradient traced by the
H\,{\sc ii} regions with nitrogen abundances determined with the direct method.

The comparison between the scatter of the residuals for the NS-calibrations 
derived here, with the scatter of the residuals for the ONS-calibrations from
\citet{pilyuginetal2010}, shows that the scatters are comparable in the case of
O/H abundances and $t$ values, but in the case of N/H abundances the scatter 
of the residual for the NS-calibrations is significantly larger than that 
for the ONS-calibrations.

\section{Conclusions}

The [O\,{\sc ii}]$\lambda$3727+$\lambda$3729 emission line is outside the
wavelength range covered by
SDSS spectra of nearby galaxies with redshift $z\la 0.023$.
Hence, one needs a calibration for deriving
abundances in H\,{\sc ii} regions such that
the calibration relations do not involve the nebular oxygen line
[O\,{\sc ii}]$\lambda$3727+$\lambda$3729.
Such relations have been derived here using the spectra of
H\,{\sc ii} regions with measured electron temperatures as calibration
datapoints.

The reliability of the NS-calibration has been verified by
comparison between a N/O--O/H diagram computed via the NS-calibration for
H\,{\sc ii} regions from SDSS spectra and that for H\,{\sc ii} regions
having abundances determined through the direct method.
We found that the NS-calibration-based abundances
follow the general trend in the N/O--O/H diagram
traced by the H\,{\sc ii} regions with the $T_e$-based abundances.

Furthermore, the NS-calibration has also been tested by comparison
between radial distributions of oxygen and nitrogen
in the disk of the well-studied spiral galaxy M~101 determined in different ways.
We found that the radial distributions of the
NS-calibration-based oxygen and nitrogen abundances agree very well with
the radial oxygen and nitrogen gradient traced by the
H\,{\sc ii} regions with abundances determined with the direct method.

We conclude that the NS-calibration provides
reliable oxygen and nitrogen abundances
over the whole range of metallicities typically found in H\,{\sc ii} regions.

Though our study was motivated by the SDSS data for nearby 
 H\,{\sc ii} regions these calibrations can be applied to any spectra 
without a reliable measurement of the nebular oxygen line [O\,{\sc ii}]$\lambda$3727+$\lambda$3729,
i.e., the spectra of many low surface brightness galaxies.
Moreover,  calibrations can be used for the abundance determination 
in any   H\,{\sc ii} region independently whether the nebular oxygen 
line [O\,{\sc ii}]$\lambda$3727+$\lambda$3729 is available or not.

\section*{Acknowledgments}

We thank the anonymous referee for helpful comments.
L.S.P. acknowledges support from the Cosmomicrophysics project of
the National Academy of Sciences of Ukraine. L.M. acknowledges support from
the Swedish Research Council.

\end{document}